\documentclass[twocolumn,preprintnumbers,amsmath,amssymb, floatfix]{revtex4}

\usepackage{graphicx}% Include figure files
\usepackage{dcolumn}% Align table columns on decimal point
\usepackage{bm}% bold math

\begin{document}

\title{Multistability in nonlinear left-handed transmission lines}

\author{David A. Powell}
\email{david.a.powell@anu.edu.au}
\author{Ilya V. Shadrivov}
\author{Yuri S. Kivshar}
\affiliation{Nonlinear Physics Center, Research School of Physical
Sciences and Engineering, Australian National University, Canberra ACT 0200, Australia}%

\begin{abstract}
Employing a nonlinear left-handed transmission line as a model
system, we demonstrate experimentally the multi-stability phenomena
predicted theoretically for microstructured left-handed
metamaterials with a nonlinear response. We show that the
bistability is associated with the period doubling which at higher
power may result in chaotic dynamics of the transmission line.
\end{abstract}

\maketitle

Left-handed metamaterials are artificial structures having negative
effective electric and magnetic parameters over some frequency range
\cite{Smith2000}.  These structures can be characterized by an
effective negative refractive index, and they support the
propagation of backward waves, where the wave fronts propagate in
the direction opposite to the energy flow defined by the Poynting
vector~\cite{Veselago1968}. This important property enables several
other interesting physical phenomena, particularly sub-wavelength
imaging~\cite{Pendry2000}.  It is also of significant interest when
combined with nonlinearity, since it enables frequency tuning,
switching between opaque and transparent states and efficient
second-harmonic reflection~\cite{ShadrivovBook}.

Left-handed transmission lines created by arrays of identical lumped
capacitors in series and shunt inductors, also support the
propagation of backward waves.  Thus such transmission lines sprovide more compact and easily studied analogues of left-handed
metamaterials, and when constructed in three-dimensional arrays they
can also form true metamaterials~\cite{Iyer2007}.  Such structures
are sometimes termed composite right-handed/left-handed transmission
lines, since the unavoidable series inductive and shunt capacitance
results in a right-handed (i.e., forward-wave) passband.

Nonlinear left-handed transmission lines have been studied for
potential device applications, and more recently for their insight
into the physics of left-handed metamaterials.  They have been shown
to exhibit mismatched harmonic generation~\cite{Kozyrev2005},
parametric generation and amplification~\cite{Kozyrev2006} and
envelope solitons~\cite{Kozyrev2007a}. In this Letter we study the
properties of a nonlinear left-handed transmission line and
demonstrate a multi-stable response in its left-handed passband,
similar to the multi-stability predicted theoretically for nonlinear
metamaterials.

%%%%%%%%%%%%%%%%%%%%%%%%%%

\begin{figure}[htb]
\includegraphics{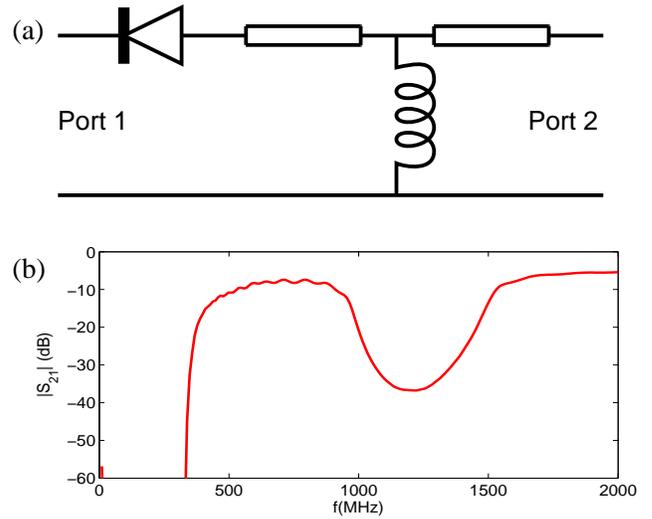}
\caption{\label{fig:unit_cell} Unit cell (a) and linear frequency response (b) of the nonlinear left
handed transmission line.}
\end{figure}

We study a micro-strip transmission line constructed on 1.6mm
copper-clad FR4.  The track sections are 3.4mm wide and 10mm long,
with 0.75mm gaps to allow the insertion of lumped capacitance and
holes to contain the shunt inductors.  The lumped nonlinear
capacitance is provided by SMV1405 hyper-abrupt junction
varactor diodes, and the inductance by 18nH chip inductors. The
equivalent circuit of each (asymmetric) unit cell is shown in figure
\ref{fig:unit_cell}(a).  This is a much simpler geometry than that
presented in \cite{Kozyrev2005}, since it has only a single diode
per unit cell and unlike the geometry presented in \cite{Gil2004}
the line is not coupled to external resonant elements.

\begin{figure}[htb]
 \includegraphics[width=\columnwidth]{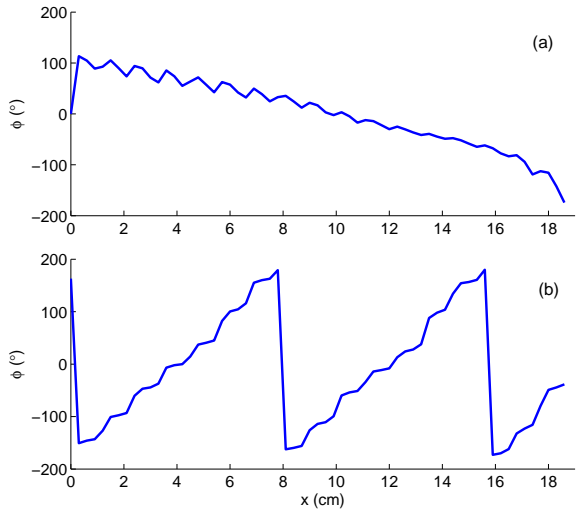}
\caption{\label{fig:phase_scan} Spatial phase distribution 1.6816GHz
in the right-handed band (a) and 607MHz in the left-handed band
(b)}
\end{figure}

The geometry and lumped element values were chosen to ensure a
significant stop-band between the right-handed and left-handed
pass-bands so that all the regimes of wave propagation in the composite
right/left-handed structure are present.  The line was 20 periods
long to ensure that end effects did not dominate the response.  As
shown in figure \ref{fig:unit_cell}(b), the left-handed passband extends
from 400-900MHz, with the right-handed band starting at approximately
1.5GHz.  The structure of the waves in the transmission line was
verified by scanning the field distribution using a probe attached
to a linear translation stage and vector network analyzer.  The
resultant phase distributions are shown in figure
\ref{fig:phase_scan}, where it can be seen that in the right-handed
band the phase decreases further away from the source, whereas in
the left-handed band it increases.

All measurements were made in a 50$\Omega$ system, however the
inherent dispersion of the left-handed transmission line means that
the characteristic impedance of the line will vary with frequency,
and this will result in reflections from the connectors. For this
reason the phase plots show some variation from linearity near the
connectors.  The non-smooth character of the curves is due to the
internal structure of the electric field distribution on the scale
of the transmission line unit cell.

In order to measure the {\em nonlinear transmission response}, a
Rohde and Schwarz SMY02 signal generator was applied to the input
port and the output port was connected to an Anritsu MS2661C
spectrum analyser.  The spectrum from 10MHz to 2GHz was captured in
501 points with a resolution bandwidth of 3MHz and video bandwidth
of 30KHz.

A sinusoidal signal of frequency 600MHz was applied, which falls
within the left-handed pass-band.  The input power was swept in
steps of 0.1 dB from 1 to 8 dBm.  Below 1dBm the system showed an
essentially linear relationship between input and output power as
well as the expected harmonic generation.

\begin{figure}[htb]
 \includegraphics[width=\columnwidth]{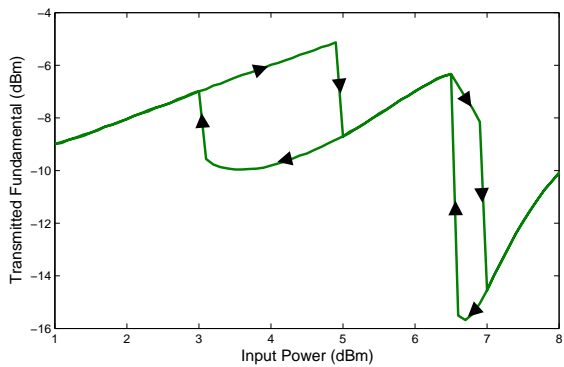}
\caption{\label{fig:fund_sweep}Output at fundamental frequency with swept input power at 600MHz, showing two regions of bistability}
\end{figure}

Due to the asymmetry of the diodes (which are all identically
oriented), the transmission responses in different directions are
not identical in the nonlinear regime.  Figure \ref{fig:fund_sweep}
shows the magnitude of the fundamental frequency as a function of
input power applied to port 1 (as defined in figure
\ref{fig:unit_cell}(a)).  It can be seen that there are two regions of
multi-stability within this power range.

\begin{figure}[htb]
 \includegraphics[width=\columnwidth]{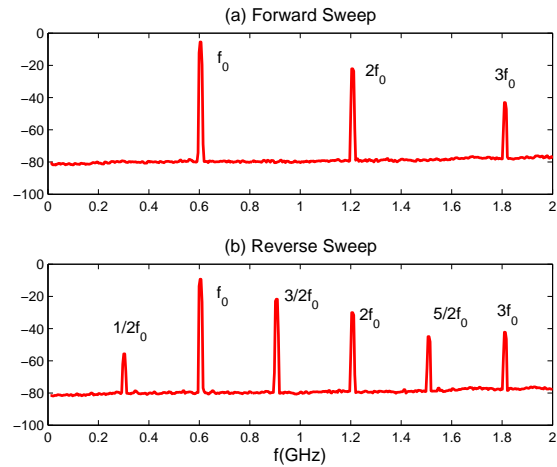}
\caption{\label{fig:spec_4p5}Output spectrum for input power of 4.5dBm for forward (a) and reverse (b) sweeps.  The reverse branch shows fractional frequency components corresponding to period doubling.}
\end{figure}

Figure \ref{fig:spec_4p5} shows the measured spectrum for an input
power of 4.5dBm swept in the forward and reverse directions.  The
forward sweep spectrum is representative of the whole branch from
the lowest measured power up to 4.9dBm.  A strong fundamental
frequency can be seen, as well as strong second and third harmonic
components.  We note that the second harmonic falls within the stop
band, however at this frequency the transmission in the stop band is
only 25dB lower than in the pass-band, thus accounting for the
relatively strong second harmonic component.

\begin{figure}[htb]
 \includegraphics[width=\columnwidth]{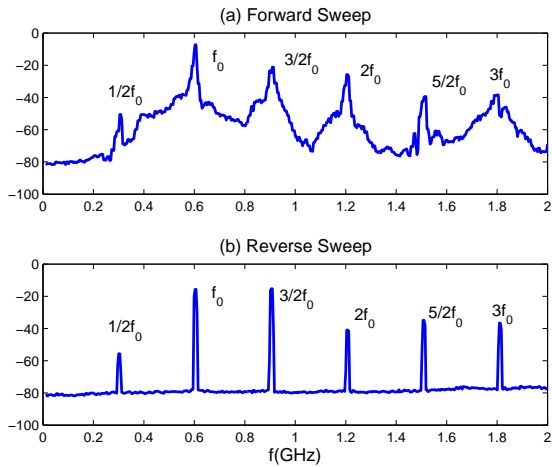}
\caption{\label{fig:spec_6p7}Output spectrum for input power of 6.7dBm for forward (a) and reverse (b) sweeps.  The forward sweep shows a quasi-continuous spectrum indicative of chaos, the reverse sweep shows period doubling.}
\end{figure}

The spectrum in the reverse sweep is representative of the lower
branch of the hysteresis curve (see Fig.~\ref{fig:fund_sweep}) extending
from 3.1dBm to the 6.5dBm (i.e. up to but not including the upper
branch of the second bistable region).  The sub-harmonic component
corresponding to a period doubling bifurcation is clearly visible in
this regime, however it falls within the lowest order cut-off band
and is strongly attenuated.  Other fractional frequency components
are much stronger, particular $3/2 f_0$, which falls within the
left-handed transmission band. The effect of multi-stability in our
nonlinear transmission line has a different physical origin 
to that in the the nonlinear left-handed metamaterials. While in metamaterials
we observe a nonlinear shift of the resonance response, in
transmission lines the transmission bands do not change, and
the existence of several dynamic states for the same input power is
related to the period doubling. The period doubling clearly
manifests itself in sub-harmonic generation for the reverse sweep
shown in Fig.~\ref{fig:spec_4p5}. It is well known that period
doubling is one of the possible routes towards chaotic behavior in a
system. Indeed, for higher powers of the signal wave, the spectrum
of the transmitted wave has quasi-continuous structure (see
Fig.~\ref{fig:spec_6p7}), which indicates chaotic behavior (see
e.g. Ref.~\cite{Schuster:Chaos}).

\begin{figure}[htb]
 \includegraphics[width=\columnwidth]{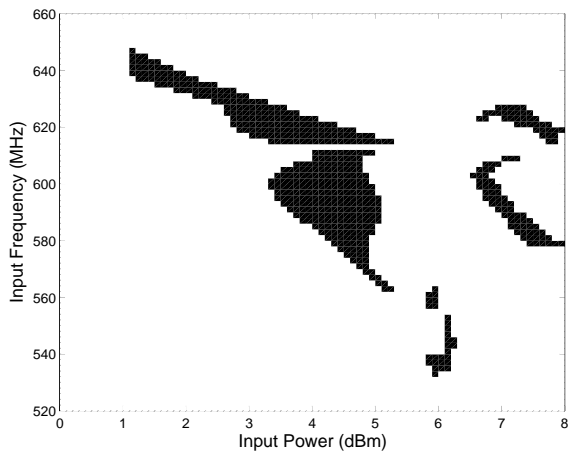}
\caption{\label{fig:map}Regions of bistability as a function of input frequency and power of a sinusoidal signal applied to port 1.}
\end{figure}

To gain a broader picture of the bistable behavior, a map of
bistable regions in frequency-power space was produced, as shown in figure
\ref{fig:map}.  This was done by performing a bidirectional power
sweep for frequencies in between 500MHz and 700MHz in 2MHz steps
(the figure has been limited to those regions where a bistable
response was found).  A threshold of 0.04dB difference in forward
and reverse power sweeps was used to eliminate the effect of
measurement noise.  This shows that several distinct regions of
bistability exist, and their frequencies suggest that the mechanism
is intimately related to the fundamental frequency being in the left
handed propagation band.  The range of frequencies also coincides
with $2f_{0}$ lying in the stop-band, $f_{0}/2$ lying within the
low-pass stop-band and $3/2f_{0}$ lying within the left-handed
passband, although it is not clear how each of these factors
contributes to the observed behavior.

For comparison purposes the fundamental response of the system was
also analyzed when the input is applied to port 2 (not shown).  The
critical points of bistability are significantly different from
those in with the input at port 1. This is due to the asymmetry in
the properties of the the diodes relative to the source, which along
with the nonlinearity produces a non-reciprocal response of the
structure.

These multi-stability results are due to a different mechanism than
that previously reported by the authors for a varactor loaded split
ring resonator (SRR) \cite{Powell2007}. For the SRR the
multi-stability was linked to the accumulation of charge across the
reverse-biased varactor diode, which resulted in the shift of the
resonance band of the resonators. In this case there is negligible
frequency shift of the band-gaps, however several dynamic states can co-exist
in the nonlinear structure at an appropriate power level.

%%%%%%%%%%%%%%%%%

In conclusion, we have experimentally studied the multi-stability
phenomenon which occurs in left-handed metamaterials with a nonlinear response. We
have employed a nonlinear left-handed transmission line as a model
system, and have demonstrated a multi-stable response in its
left-handed passband, similar to the multi-stability predicted
theoretically for nonlinear metamaterials. We have shown that the
bistability is associated with period doubling which may result in chaotic dynamics at higher
power.

The authors thank Dr. Thomas Stemler for helpful discussions and
acknowledge support from the Australian Research Council through the
Discovery Project.

\end{document}